\documentclass[pdflatex,sn-nature,Numbered]{sn-jnl}

\usepackage{lmodern}   
\usepackage{graphicx}
\usepackage{multirow}
\usepackage{amsmath,amssymb,amsfonts}
\usepackage{amsthm}
\usepackage[title]{appendix}
\usepackage{xcolor}
\usepackage{textcomp}
\usepackage{manyfoot}
\usepackage{booktabs}
\usepackage{algorithm}
\usepackage{algorithmicx}
\usepackage{algpseudocode}
\usepackage{listings}
\usepackage{hyperref}
\hypersetup{colorlinks=true, urlcolor=blue, citecolor=blue, linkcolor=blue}

\definecolor{cPurple}{HTML}{9238E0}  
\definecolor{cBrown}{HTML}{E09238}   
\definecolor{cGreen}{HTML}{38E092}   
\definecolor{cPurpleLt}{HTML}{C197E8}  
\definecolor{cBrownLt}{HTML}{E8C197}   
\definecolor{cGreenLt}{HTML}{C7E9C0}   
\definecolor{cTurqLt}{HTML}{97E7E8}    
\definecolor{cTurq}{HTML}{38DAE0}      
\definecolor{cBlueLt}{HTML}{9897E8}    
\definecolor{cBlue}{HTML}{3E38E0}      
\definecolor{cGreyLt}{HTML}{DBDBDB}    
\definecolor{cGrey}{HTML}{707070}      

\theoremstyle{definition}

\newtheorem{definition}{Definition}[section]

\begin{document}

\title[A global mobile network coverage raster, 1999--2030, at 1\,km] {A global mobile network coverage raster product at 1\,km resolution, 1999--2030}

\author*[1]{\fnm{Till} \sur{Koebe}}\email{till.koebe@uni-saarland.de}  \author[1]{\fnm{Theophilus} \sur{Aidoo}}
\author[1]{\fnm{Ali} \sur{El Chami}} 
\author[1]{\fnm{Ali} \sur{Kanso}} 
\author[1]{\fnm{Akansh} \sur{Maurya}} 
\author[2]{\fnm{Purushottam} \sur{Sharma}} 
\author[1]{\fnm{Ingmar} \sur{Weber}}
\author[3]{\fnm{Ridhi} \sur{Kashyap}} 

\affil[1]{\orgname{Saarland Informatics Campus}} \affil[2]{\orgname{IIT Delhi}} \affil[3]{\orgname{Oxford University}}

\abstract{Where a mobile signal is available shapes who can work, learn, bank, seek health care and respond to crises in the digital age, yet no globally consistent, sub-national record of mobile network coverage exists. We present such a record: annual 1\,km maps of the probability of 2G, 3G and 4G coverage for 214 countries and territories for the years 1999 to 2030. The maps are produced by three independent models: a calibrated machine-learning model, a techno-economic simulator of network build-out, and a spatial deep-learning model. The three estimates are then combined, per country and technology and in proportion to their measured accuracy, into a single best estimate with per-pixel 90\,\% uncertainty bands; all four layers are released as part of the dataset. Because mobile roll-out closely follows a country's socio-economic conditions (population distribution, electrification, physical infrastructure), the models are grounded in existing geospatial data and tuned on 2{,}409 quality-screened operator-reported coverage maps, which are available up to 2020. For 2021--2024 the maps are predicted from recent geospatial data alone; for 2025--2030 they are extrapolated from demographic and infrastructure projections. On countries held out during training, the machine-learning model attains AUC 0.89--0.92. Baseline comparisons and the combined product's external validation are reported in Technical Validation. The dataset supports mapping the global digital divide, linking connectivity to household-survey outcomes, and humanitarian and infrastructure planning.}


\maketitle

\section*{Background \& Summary}\label{sec:background}

Access to a mobile signal is the prerequisite for participation in the modern digital economy. While the country-year prevalence of mobile subscriptions is well documented \cite{itu_facts_figures_2024}, sub-national distributions of \emph{network coverage}, i.e. the supply side, where service is physically available, are much less consistently observed. The Global System for Mobile Communications Association (GSMA), through its mapping partner Collins Bartholomew, has published the Mobile Coverage Explorer (MCE) annually since 2009. Each MCE release aggregates operator-supplied coverage polygons by technology (2G, 3G, 4G, 5G) and country, producing global rasters at approximately 265\,m resolution \cite{collins_bartholomew_mce}.

The MCE dataset has structural limitations that have prevented its use as a reliable cross-national or longitudinal coverage source. First, \emph{not every operator reports every year}: in any given MCE release, the proportion of countries whose major operator updated within the preceding two years has historically been below 50\,\%. Second, even where a country shows a recent update, the accompanying metadata identifies \emph{which} operator updated its map but not whether that operator is \emph{representative} of national coverage. For population coverage this matters greatly: an update from a country-wide incumbent whose footprint comes closest to the country's actual served extent is far more informative than one from a small or specialised provider whose map covers only a niche (often urban) and understates rural connectivity, and the metadata does not flag which of the two it is. Classifying, per country, which operators constitute \emph{major} operators (Definition~\ref{def:major}) is therefore a necessary step, and a contribution of this work. Third, how each operator generates its coverage map is a black box: dozens of different prediction and measurement approaches are in use and can produce substantially different footprints for the same network \cite{warwick2022}. We acknowledge this source heterogeneity but do not attempt to correct for it. Naive use of MCE thus biases sub-national analyses toward whichever operators happen to maintain their submissions in a given year.

Other efforts to map mobile coverage at scale draw on crowdsourced, app-, or SDK-based measurements rather than operator-reported maps. App-derived signal traces underpin large open cell databases such as OpenCellID \cite{opencellid} and the now-retired Mozilla Location Service, and reconstructing cell footprints from such crowdsourced received-signal-strength data is feasible but error-prone, with high variance and systematic gaps where few volunteers contribute \cite{li2017crowdrss,fida2019uncovering}. Commercial SDK panels such as Ookla Speedtest \cite{ookla_speedtest} and Opensignal supply the measurements behind many regulators' and platforms' coverage and performance benchmarks \cite{midoglu2017closerlook,ofcom2025mobilematters}, yet their spatial sample is conditioned by where users run tests, producing demographic and geographic bias that under-represents exactly the access-gap areas of greatest policy interest unless explicitly corrected \cite{lee2023disparity,sharma2024spatial}. Platform products such as Meta's Data-for-Good settlement and population layers and its Disaster Maps connectivity insights derive population and connectivity surfaces from app usage and satellite imagery, but are typically population- or crisis-anchored and gated behind humanitarian-partner access rather than offered as open, persistent coverage maps \cite{tiecke2017mapping,maas2019disastermaps}. Modelling approaches that fuse operator call-detail records, radio-propagation models and remote sensing can reconstruct footprints at scale, but require privileged access to a single operator's data \cite{koebe2020bettercoverage}. Authoritative ITU and GSMA statistics, finally, aggregate operator- and regulator-reported figures and report \emph{population covered} rather than a spatially explicit, openly redistributable footprint \cite{itu_facts_figures_2024,gsma2022somic}. Reliable-operator-map coverage thus fills a distinct niche: a globally consistent, spatially explicit, openly available footprint that avoids both the app-penetration bias of crowdsourced datasets and the access restrictions of platform- and operator-derived products.

Mobile connectivity is not only a development outcome to be tracked; it is a major driver of societal change, reshaping how people work, trade, learn and communicate, yet granular information on it is patchy. What ultimately produces these impacts is not the infrastructure itself but its \emph{use}: the same connection can carry passive consumption or a new business. Use, however, is very hard to observe at scale, so research falls back on more measurable proxies such as device ownership, subscriptions, or self-reported mobile internet access. All of these presuppose one thing that is itself poorly mapped: that a usable signal is available at all. A consistent sub-national record of coverage, the \emph{precondition} for use, is therefore foundational for studying the societal impact of mobile connectivity, from the demand side proxies above down to where networks reach in the first place. Methodologically, inferring fine-grained infrastructural conditions where direct observation is sparse has been addressed by combining Earth observation with machine learning. One lineage predicts latent socioeconomic surfaces such as local economic well-being from satellite imagery and mobile-phone metadata \cite{jean2016,blumenstock2015,steele2017,yeh2020,burke2021}, which establishes that the globally available raster layers we use (i.e. population, built-up surface, roads and nighttime lights) are predictive of where networks are built. A closer analogy for the \emph{infrastructure} itself is the prediction of electrification rollout: nighttime lights are a ground-validated signal of village-level access \cite{min2013detection,falchetta2019}, and Gridfinder reconstructs the global power grid by combining them with roads and a least-cost routing algorithm to a structural, cost-informed model of where lines must run instead of a statistical correlate of prosperity \cite{arderne2020predictive}. This is precisely the logic of our structural simulator (Section~\ref{sec:trackB}): like grid lines, mobile towers are sited to extend service from existing infrastructure under explicit cost constraints, mirroring least-cost electrification planning \cite{mentis2017lighting,korkovelos2019role}, and the two infrastructures co-evolve \cite{salat2021analysing}.

This motivates treating coverage modeling as a supervised spatial-prediction problem on the same class of covariate stack, anchored to a carefully filtered observational label set rather than to the crowdsourced signals discussed above.

We address these limitations by (i) restricting supervised use of MCE to a transparently-defined \emph{reliable} label set (Definitions~\ref{def:major}--\ref{def:reliable}), and (ii) filling all other country-tech-year combinations with predictions from three independent modelling tracks. The reliable set comprises 2{,}488 accepted rows, resolving to 2{,}409 unique country-technology-year triples across 207 of the 214 reference-grid countries (those with at least one reliable triple; the remaining seven are listed in Data Records): 1{,}058 grounded in modern MCE operator metadata (publication years 2010--2021) and a further 1{,}430 recovered from the pre-2010 MCE legacy archive (coverage years 2000--2009) by a documented substitute reliability filter (Section~\ref{sec:reliability}), with reliable label years spanning 2000--2020. We model 2G, 3G and 4G; 5G appears in recent MCE releases but yields too few reliable triples across countries and years to support the supervised filter.

Where OpenCellID's volunteer-collected tower counts are known to under-represent legitimate major operators, e.g.\ China Telecom (real subscriber share $\sim$24\,\% but only 3.9\,\% of OpenCellID tower count vs.\ China Mobile), Iran (sanctions limit OpenCellID volunteer-app distribution), Vodafone Idea in India, KDDI and SoftBank in Japan, we override the 75\,\% threshold with a curated per-country additional-majors list, sourced from national regulators (TRAI, MIIT, NCC, BTRC, KCC, IFT, CITC, NTRA, PTA) and major industry trackers (GSMA Intelligence, Statista).

Three tracks each ingest the same covariate stack and produce a per-pixel-year-tech coverage probability. Track~A applies a gradient- boosted decision-tree classifier with monotonicity constraints and isotonic post-hoc calibration. Track~B is a techno-economic simulator that places towers greedily by predicted profit (population $\times$ average revenue per user (ARPU) within an antenna footprint, minus tower construction and backhaul cost), with country-specific parameters calibrated by the Covariance Matrix Adaptation Evolution Strategy (CMA-ES) against the reliable label set. Track~C is a pix2pix-style spatial deep-learning model that treats coverage prediction as image translation conditional on the multi-channel covariate stack. We release all three layers and a \emph{best-estimate} composite that combines them per pixel, weighting each track by its measured accuracy on the reliable labels for that country and technology.

No single modelling paradigm dominates this problem, which is why we deploy three methodologically independent tracks rather than one. Track~A is a statistical learner that excels at exploiting the tabular covariate stack but is, beyond its buffer features, largely blind to spatial context. Track~B is a techno-economic simulator in the tradition of bottom-up telecom-infrastructure costing models \cite{oughton2019}, which encode the engineering and economic logic of network rollout and yield interpretable per-country deployment-friction parameters, at the cost of strong structural assumptions. Track~C is a spatial deep-learning model based on conditional image translation \cite{isola2017,ronneberger2015} that learns coverage morphology directly from the raster context but depends on dense labels. Because the three err in different, weakly correlated ways, a calibrated combination is more robust than any one in isolation. Each track's probabilistic output is post-hoc calibrated \cite{niculescu2005}; the combined estimate is a validation-weighted mean of the three, and its released 90\,\% interval is obtained by split-conformal calibration against held-out reliable labels (see Section \ref{sec:methods}).

\begin{figure}[t]
\centering
\includegraphics[width=1\linewidth]{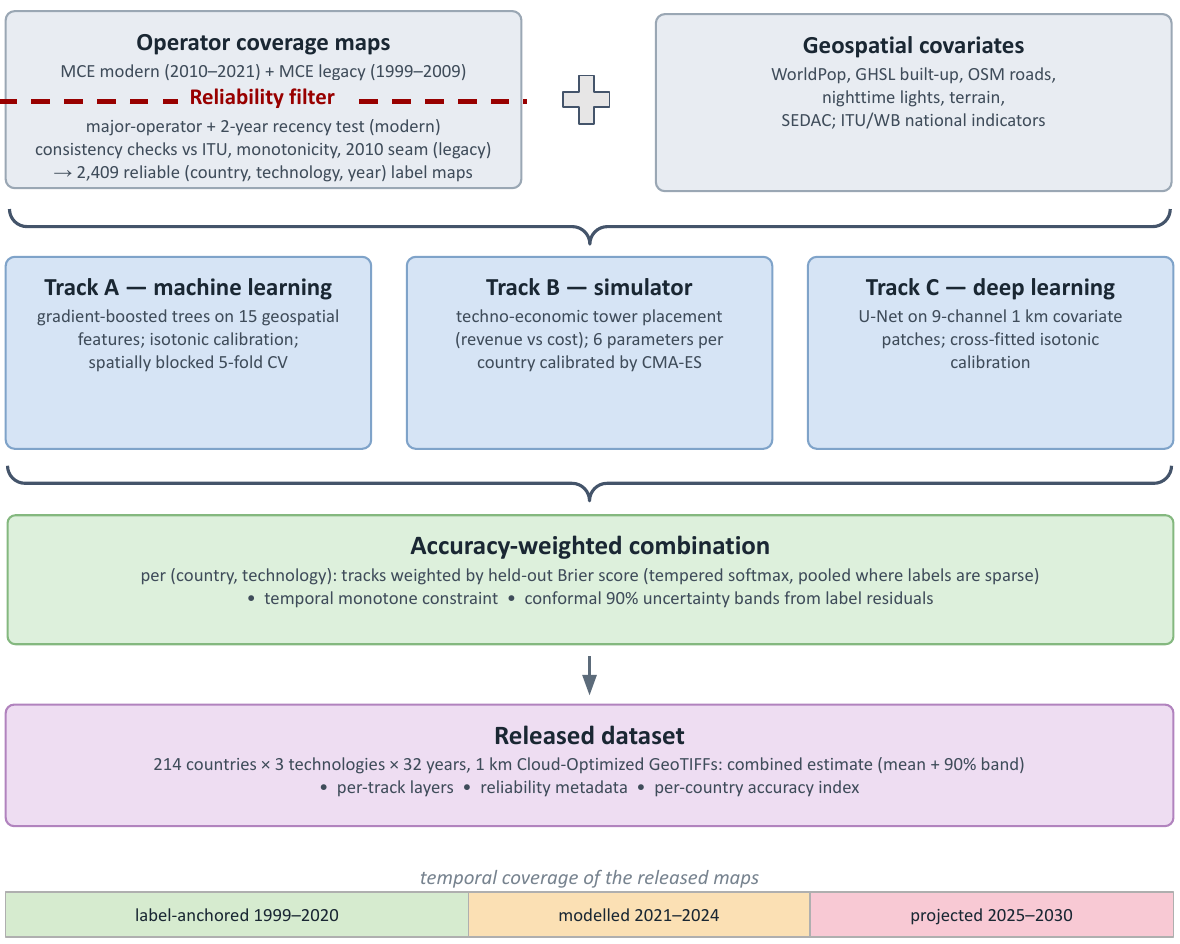}
\caption{Dataset-construction pipeline. Operator-reported coverage maps are
screened by the reliability filter into 2{,}409 label maps, which train and
calibrate three independent estimation tracks; per (country, technology) the
tracks are combined in proportion to their held-out accuracy, constrained to
be temporally monotone, and wrapped in conformal 90\,\% uncertainty bands.
The ribbon indicates the temporal regimes of the released maps:
label-anchored (1999--2020), modelled (2021--2024) and projected
(2025--2030).}
\label{fig:pipeline}
\end{figure}

The product is a collection of cloud-optimised GeoTIFFs on a common 1\,km grid (EPSG:4326), one per (country, technology, year) for 214 countries, three technologies (2G/3G/4G) and annual time steps 1999--2030, in four estimation layers: the combined best estimate (weighted mean with a calibrated 90\,\% interval) plus the three per-track point estimates (see Figure \ref{fig:pipeline}). 

To our knowledge no comparable open product exists: the International Telecommunication Union's (ITU) DataHub provides only country-level aggregates \cite{itu_datahub}, GSMA Intelligence is paywalled and at operator level, and OpenSignal/Ookla \cite{ookla_speedtest} cover network quality rather than coverage extent. The full dataset, model artefacts, and source code are available under CC-BY-4.0 (rasters) and BSD-3-Clause (code).

\section*{Methods}\label{sec:methods}

\subsection*{Reference grid construction}\label{sec:grid}

The reference grid is a per-country 1\,km (nominal) grid in EPSG:4326, derived deterministically from each country's WorldPop UN-adjusted 1\,km population raster \cite{worldpop_unadj}: one axis-aligned cell per valid (non-nodata) raster pixel, via the raster's affine transform. Cells with zero population on valid land are retained, so the grid spans each country's entire land surface, and every covariate, label and prediction in the product is co-registered on it.

\subsection*{Reliability filter}\label{sec:reliability}

\begin{definition}[Major operator]\label{def:major}
For country $c$, a major operator is either: (i) any operator whose OpenCellID 2G and above (2G+) tower count is at least 75\,\% of the country-maximum operator's tower count; or (ii) an operator on the curated additional- majors override list for $c$ (see Appendix \ref{tab:overrides}).
\end{definition}

\begin{definition}[Reliable coverage triple]\label{def:reliable}
A triple $(c, t, y)$ where $c$ is a country, $t \in \{2\mathrm{G}, 3\mathrm{G}, 4\mathrm{G}\}$ a technology and $y$ a coverage-data year is \emph{reliable} if at least one major operator of $c$ reported a coverage update for technology $t$ in year $y$ with $y' - 2 \le y \le y'$, where $y'$ is the MCE publication year, i.e.\ the update is no more than two years older than the publication it appears in.
\end{definition}

Major operators are identified from the OpenCellID 2024-01-20 tower dump \cite{opencellid}, with operator names matched to MCE entries. For 16 countries where OpenCellID tower counts demonstrably under-represent real major operators, a researched override list adds those operators explicitly; the full list, with one row per country, is given in Appendix~\ref{tab:overrides}.

MCE labels are attached per cell by one of two samplers. The default computes the covered \emph{fraction} of each 1\,km cell over the (higher-resolution) MCE raster and binarises it at $0.5$ (majority rule); for the eight continental-scale countries whose grids are built by the chunked builder (pixels $>4$\,M), the label is instead the MCE value sampled at the cell centroid; the two rules agree away from coverage frontiers and differ only at partially-covered edge cells. MCE encodes values 1 and 2 as ``covered'' (different signal classes), value 3 as ``not covered''. The raster's nodata metadata is also set to 3; we explicitly map values 1--2 to covered, value 3 to not covered, and treat out-of-raster sampling positions and missing rasters as NaN (no label), to avoid the common pitfall of treating ``not covered'' as missing data.

Applying this recency filter to the 2010--2021 MCE releases, of the 1{,}203 candidate (country, technology, publication) triples reported by a major operator, 1{,}058 are accepted (87.9\,\%; Table~\ref{tab:modern}), spanning coverage years 2008--2020 across 168 countries (the two-year recency window of 2010/2011 publications reaches back to 2008--2009. Where a modern and a legacy map compete for the same (country, technology, year), the modern map takes precedence, and among competing modern publications the earliest wins). Acceptance rises with technology recency (80.9\,\% for 2G but 98.1\,\% for 4G).

\begin{table}[!ht]
\centering
\small
\begin{tabular}{lcccc}
\toprule
\textbf{Tech} & \textbf{Candidate} & \textbf{Accepted} & \textbf{Rate} &
\textbf{Recency-} \\
 & \textbf{triples} & & & \textbf{dropped} \\ \midrule
2G  & 554     & 448     & 80.9\,\% & 106 \\
3G  & 441     & 406     & 92.1\,\% & 35  \\
4G  & 208     & 204     & 98.1\,\% & 4   \\
All & 1{,}203 & 1{,}058 & 87.9\,\% & 145 \\
\bottomrule
\end{tabular}
\caption{Acceptance of modern MCE triples (publication years
2010--2021) by the operator-recency reliability filter (Section~\ref{sec:reliability}). ``Candidate triples'' are the deduplicated (country, technology, publication) tech-cells reported by a \emph{major} operator (Definition~\ref{def:major}): from the 6{,}163 MCE operator entries in this window, 3{,}012 match a major operator (377 majors across all countries, including the 16-country override list), collapsing to 1{,}203 candidate triples after deduplication to the earliest publication. ``Recency-dropped'' triples failed the $y' - 2 \le y \le y'$ update window (Definition~\ref{def:reliable}): the operator's last update for that technology was more than two years older than the publication it appeared in. The 1{,}058 accepted triples (522 country-publication rows, 168 countries) constitute the modern half of the reliable label set. This is the 2010-onward counterpart of the pre-2010 legacy acceptance in Table~\ref{tab:legacy}.}
\label{tab:modern}
\end{table}

\paragraph{The pre-2010 legacy archive and its substitute filter.}
The pre-2010 MCE legacy archive (1999--2009) is distributed as raw ESRI shapefiles in a format incompatible with the modern MCE rasters. As part of this work we harmonised it: 1{,}956 per-(country, year, technology) rasters re-encoded to the modern MCE convention and aligned to each country's MCE spatial template, so that legacy and modern maps can be sampled identically. This harmonisation is what makes the 1999--2009 extension of the dataset possible. The legacy archive, however, carries no per-operator metadata, so the reliability filter above cannot be applied to it. To admit the legacy maps without the operator-recency test we introduce a substitute filter that accepts a legacy triple $(c, t, y)$ if and only if it passes \emph{all applicable} consistency checks, i.e. a check whose input data are unavailable is skipped. A triple for which \emph{no} check is applicable is rejected as unverifiable (a conservative default that removes 267 of the 1{,}947 candidates, including all 108 from 1999, which is why the accepted set begins in 2000 although the archive starts in 1999). The checks are: (i) \emph{ITU agreement}, the population-weighted coverage share computed from the legacy raster lies within $\pm 25$ percentage points of ITU's reported $(c, t, y)$ coverage. Skipped where ITU records no value (common before $\sim$2005); (ii) \emph{temporal monotonicity} meaning the covered area does not shrink anomalously year-on-year, $\mathrm{area}(y) \ge 0.95 \times \mathrm{area}(y-1)$ per $(c, t)$. Skipped for the first archived year; and (iii) \emph{2010-boundary consistency}, i.e. the legacy 2009 footprint agrees with the modern MCE 2010 map, $\mathrm{area}(\mathrm{legacy}_{2009} \cap \mathrm{mce}_{2010}) / \mathrm{area}(\mathrm{legacy}_{2009}) \ge 0.8$, applicable only at the 2009 seam. The monotonicity check flags anomalous shrinkage but, by design, lets stagnant footprints pass: a map that is simply not being updated is indistinguishable from a genuinely static network. We accept this because flat coverage is the expected state for the mature 2G networks that dominate the archive and, where those checks are applicable, the ITU-agreement and 2010-boundary tests provide partial corroboration; it remains a limitation for technologies in active rollout. Accepted rows enter \texttt{reliable\_coverages.csv} in the same schema as the modern set but with \texttt{pub\_year} null and a \texttt{source} column set to \texttt{legacy} (vs \texttt{mce}).

Of the 1{,}947 candidate legacy triples whose country falls within the 214-country reference grid (2G and 3G only, 4G postdates the archive), 1{,}430 are accepted (73.4\,\%; Table~\ref{tab:legacy}), spanning coverage years 2000--2009. Of these, 664 (46\,\%) had an independent ITU country aggregate available; the remaining 766 (54\,\%) carry no ITU anchor and rest on the temporal-monotonicity and 2010-boundary checks alone. Per-check outcomes for every candidate triple are released in \texttt{reliable\_legacy\_report.csv}. Users who want only the ITU-based legacy labels can keep the 664 accepted triples that passed the ITU-agreement check.

The resulting label set is tilted toward legacy 2G: of the 2{,}488 reliable triples (2{,}409 distinct (country, technology, year) maps, since a triple may appear in more than one MCE publication), 1{,}325 are legacy 2G, against 448 modern 2G, 511 (modern + legacy) 3G and 204 (modern-only) 4G. This asymmetry is expected: 2G was the mature, dominant technology of 2000--2009, and its stable footprints pass the consistency checks easily, thus appears more permissive than the modern operator-recency requirements. Two design choices address the imbalance: the per-country sample weighting and pixel cap (Section~\ref{sec:trackA}) prevent any one country-era from dominating training, and the temporal hold-out (Technical Validation) provides an era-split check, since its training window ($\le 2015$) is legacy-dominated while its test window ($\ge 2016$) is purely modern.

\begin{table}[!ht]
\centering
\small
\begin{tabular}{lccccc}
\toprule
\textbf{Tech} & \textbf{Candidate} & \textbf{Accepted} & \textbf{Rate} &
\textbf{ITU-checked} & \textbf{Consistency-only} \\
 & \textbf{triples} & & & \textbf{(median $|\Delta|$)} & \textbf{(mono+bndry)} \\ \midrule
2G  & 1{,}754 & 1{,}325 & 75.5\,\% & 781            & 544 \\
3G  & 193     & 105     & 54.4\,\% & 48             & 57  \\
All & 1{,}947 & 1{,}430 & 73.4\,\% & 664 (1.8\,pp)  & 766 \\
\bottomrule
\end{tabular}
\caption{Acceptance of pre-2010 legacy MCE triples by the substitute
reliability filter (Section~\ref{sec:reliability}). ``ITU-checked'' triples had an ITU country aggregate available and passed the $\pm 25$\,pp agreement test (observed median absolute difference 1.8\,pp); ``consistency-only'' triples had no ITU value for that (country,~year) and were admitted on temporal monotonicity and 2010-boundary overlap alone. No legacy 4G exists (4G postdates the archive). The full per-check overview is released as \texttt{reliable\_legacy\_report.csv}.}
\label{tab:legacy}
\end{table}

\subsection*{Covariate layers}\label{sec:covariates}

We choose covariates to capture the three forces that govern where a network is built: \emph{demand} (where people and settlements are), \emph{economic value} (what that demand is worth to an operator), and \emph{deployment feasibility} (how hard and costly a site is to build and connect). Observed and projected population and built-up surface locate and quantify demand, while the GHSL settlement model distinguish urban cores, rural settlement and unpopulated land that no network targets. Nighttime lights, the gridded deprivation index and the buffer-aggregated revenue-potential features proxy ability-to-pay, and hence the commercial incentive to deploy that also drives Track~B's economics. Terrain elevation and ruggedness stand in for radio line-of-sight and tower-siting cost, the road network for both site access and backhaul corridors, and the two World Bank telecom indicators anchor each country-year to its overall market maturity (the prediction year carries the temporal rollout). The binding constraint on this set is not predictive power but \emph{global consistency}: every layer must be available for all 214 countries and resolvable across the 1999--2030 horizon, i.e. time-varying where a series exists (population, built-up surface, nighttime lights), held flat where only a static layer is available (terrain, deprivation), which excludes many otherwise-useful sources (for a full list of covariates used see Table~\ref{tab:covariates}).

\begin{table}[!ht]
\centering
\footnotesize \setlength{\tabcolsep}{4pt}
\resizebox{\textwidth}{!}{\begin{tabular}{@{}l l l l l@{}}
\toprule
\textbf{Layer} & \textbf{Source} & \textbf{Years} & \textbf{Resolution} & \textbf{Licence} \\ \midrule
\multicolumn{5}{@{}l}{\textit{Global feature set}}\\
Population (observed) & WorldPop UN-adj 1\,km PPP \cite{worldpop_unadj}     & 2000--2020       & 1\,km     & CC-BY-4.0 \\
Population (projected) & WorldPop Global-2 R2024B, national totals only \cite{worldpop_r2024b} & 2021--2030       & national  & CC-BY-4.0 \\
Built-up surface & GHSL R2023A BUILT-S \cite{ghsl_r2023a}                   & 1975--2030 (5y) & 30\,arcsec & CC-BY-4.0 \\
Settlement model & GHSL R2023A SMOD \cite{ghsl_r2023a}                      & 1975--2030 (5y) & 1\,km      & CC-BY-4.0 \\
Roads & OSM history, drive-net classes \cite{osm_drive}                     & 2000--present    & vector    & ODbL \\
Nighttime lights & Li et al.\ harmonised \cite{li2024ntl}                   & 1992--2024       & 30\,arcsec & CC-BY-4.0 \\
Terrain & NOAA ETOPO 2022 30s \cite{etopo2022}                              & static           & 30\,arcsec & PD \\
TRI \cite{riley1999tri} & ETOPO-derived                                    & static           & 30\,arcsec & PD \\
Country-year telecom & WB WDI ICT indicators \cite{wb_wdi}                  & 2010--present    & country    & CC-BY-4.0 \\
Country-year coverage & ITU DataHub \cite{itu_datahub}                      & 2010--2024       & country    & CC-BY-NC-SA-3.0-IGO \\
Static deprivation & SEDAC GRDI v1 \cite{sedac_grdi}                        & static           & 1\,km      & CC-BY-4.0 \\
\bottomrule
\end{tabular}}
\caption{Input covariate layers. Every feature in the global set is
computable for each of the 214 countries in the product. GHSL BUILT-S is linearly interpolated to annual values between its 5-year epochs; SMOD is held flat at the nearest epoch. Nighttime lights span 1992--2024; values are held flat at 2024 across the 2025--2030 projection horizon. The two World Bank telecom indicators (mobile subscriptions, internet use), observed 1999--2024, are likewise completed over the full range: internal gaps are linearly interpolated and the series are held at the nearest observed value beyond their span (last observed 2024), so forecast-year inference receives in-distribution national covariates rather than missing values.}
\label{tab:covariates}
\end{table}

Indeed, the most informative covariates are precisely those that do not exist as open, global, time-resolved layers. Operator-internal data such as base-station locations and configuration, backhaul and fibre routes, spectrum holdings, and measured traffic from call-detail records would largely determine coverage but are confidential; where they are crowdsourced (e.g.\ OpenCellID tower dumps \cite{opencellid}) they are incomplete and biased toward where users run apps, and modelling from a single operator's records does not generalise \cite{koebe2020bettercoverage}. A consistent global, time-varying electricity-grid layer would be a strong predictor given the close coupling of the two infrastructures, but global grid maps are \emph{reconstructed} rather than observed and lack a reliable annual series \cite{arderne2020predictive}. Other relevant layers \emph{are} available but are kept out of the global model: high-resolution building-footprint datasets exist for recent years but not as a 1999--2030 series, so GHSL built-up surface is preferred for temporal consistency; and Meta's Relative Wealth Index \cite{chi2022rwi} and Falchetta's electrification surface \cite{falchetta2019} are strong economic proxies but cover only low- and middle-income countries and sub-Saharan Africa, respectively.

\subsubsection*{Buffer-aggregated economic features}

For each pixel we compute three buffer aggregates within a 10\,km radius: the population sum, the population $\times$ ARPU sum (where ARPU is country-year-known), and the built-up surface sum. These ``revenue potential'' buffer aggregates mirror the logic of Track~B (Section~\ref{sec:trackB}) and are computed on the WorldPop raster.

\subsection*{Track A: gradient-boosted decision trees}\label{sec:trackA}

We train one LightGBM classifier per technology \cite{ke2017lightgbm}. The 15-feature set comprises the per-pixel layers from Table~\ref{tab:covariates}, the three buffer aggregates (Section~\ref{sec:covariates}), the prediction \texttt{year}, and the two World Bank telecom indicators (mobile subscriptions and internet-user share per 100). Two of the fifteen are categorical (the GHSL SMOD settlement class and a country code). Monotonicity constraints are enforced on features whose causal sign is known a priori such as population, the population- and built-up buffer aggregates, built-up surface, nighttime lights, and mobile and internet penetration are constrained $+1$; terrain ruggedness and elevation $-1$, so that the learned response cannot, for example, fall as population rises. The constraints inject domain knowledge that the labels alone may not pin down: reliable labels are sparse and spatially clustered, so an unconstrained tree can latch onto locally spurious reversals (lower predicted coverage at higher population) that do not generalise. Fixing the causal sign keeps predictions physically plausible and steadies extrapolation to country-years far from the labelled data.

Country sizes span more than four orders of magnitude (Andorra $\sim$$10^3$ pixels, Russia $\sim$$10^7$), which we address with two separate mechanisms. For computational tractability, each country contributes at most $10^6$ pixels per year, drawn as a deterministic uniform random subsample. For statistical balance, rows are weighted by $w_i = 1/\sqrt{n_c}$ (with $n_c$ country $c$'s row count), so that large and small countries contribute comparably to the training objective. The boosters use a binary log-loss objective with conservative regularisation settings, fixed by design rather than tuned per run, to curb overfitting to country-specific patterns on the lower-$N$ technologies, most acutely 4G (all settings in Table~\ref{tab:hyperparams}, Appendix~\ref{app:specs}). We calibrate post-hoc with isotonic regression \cite{niculescu2005}, i.e. a monotonic, non-parametric rescaling of the raw model scores into probabilities, fit so that among cells assigned a given score the observed coverage rate matches it (of the cells scored near $0.7$, about $70\%$ are in fact covered), on a held-out split that the reported metrics never see. Generalisation is estimated by 5-fold spatial-block cross-validation over countries (Section~\ref{sec:internalcv}), each fold a disjoint round-robin country assignment.

\subsection*{Track B: structural simulator}\label{sec:trackB}

Each pixel is a candidate omnidirectional antenna site. Expected annual revenue of a site is the population within its antenna footprint, valued at the country-year ARPU (spatially disaggregated, see below) and penetration; its cost is either a cheap upgrade of an existing lower-technology tower or a greenfield build plus backhaul. For technology $t$, country $c$, year $y$, pixel $i$:
\begin{align}
R_{i,y,t} &= \sum_{j \in B_t(i)} \mathrm{pop}_{j,y} \cdot
            f(\mathrm{SEDAC}_j) \cdot
            \mathrm{ARPU}_{c,y,t} \cdot
            \mathrm{pen}_{c,y,t} \cdot 12 \cdot \alpha_{c,t}, \\
C_{i,y,t} &= \begin{cases}
   C^{\text{up}}_{c,t} & \text{if a tower of tech} < t \text{ at } i, \\
   C^{\text{green}}_c + \min(c^{\text{fibre}}_c \cdot d_i, c^{\text{mw}}_c) & \text{else,}
\end{cases}
\end{align}
where $B_t(i)$ is the antenna footprint (fixed radius $r_t \in \{12,7,5\}$\,km for 2G/3G/4G; a single per-technology constant everywhere: density-dependent cell sizes are not modelled), $f(\mathrm{SEDAC}_j)$ disaggregates the country-level ARPU to the pixel level via SEDAC GRDI deprivation ($f(\mathrm{SEDAC}_j) = e^{-\beta z_j} / \mathbb{E}[e^{-\beta z}]$ with $z_j$ the within-country population-weighted deprivation z-score and $\beta{=}0.3$ fixed, preserving the country mean: less-deprived pixels carry proportionally more revenue), and $d_i$ is the great-circle distance from $i$ to the nearest existing tower as of $y-1$. The annual build budget is endogenous, $\mathrm{Budget}_{c,y} = \mathrm{ARPU}_{c,y} \cdot \mathrm{subscribers}_{c,y} \cdot 12 \cdot \mathrm{capex\_share}_c$ from ITU series \cite{itu_datahub}, bound by the technology availability gate of Section~\ref{sec:gate} (a technology cannot deploy in $c$ before it was commercially available there: the certified launch year where one exists, and the ITU first-positive-coverage year otherwise) and modulated by a tech-specific \emph{maturity ramp}, i.e. a smooth factor that scales deployment up as the technology matures in a market. For each (country, tech) with at least three reliable triples we fit six free parameters by CMA-ES \cite{hansen2003}, a derivative-free optimiser that iteratively adapts a sampling distribution to minimise a loss, here a \emph{Dice} loss, which measures the spatial overlap between the predicted and observed covered areas ($1$ = perfect overlap, $0$ = none), with log-normal priors anchored to published mobile-infrastructure cost estimates (tower build, upgrade and backhaul costs) \cite{oughton2019}; (country, technology) cells with fewer triples inherit the median posterior of their World Bank income group. The full prior, constant and optimiser specification is given in Appendix~\ref{app:specs} (Table~\ref{tab:hyperparams}).

Track~B rests on strong structural assumptions, stated here in full: (i) operators deploy \emph{greedily} to maximise predicted profit, building towers in descending profit order until the annual budget is exhausted; (ii) each site is a single \emph{omnidirectional} antenna and the model represents coverage extent, not capacity or congestion; (iii) demand at a site is proportional to population within the footprint times country-year ARPU and penetration; ARPU is spatially disaggregated within the country along the SEDAC deprivation gradient (a fixed $\beta{=}0.3$ assumption, not calibrated), while penetration is taken as country-uniform; (iv) the annual build budget is a calibrated share of sector revenue, taken endogenously from ITU subscriber and ARPU series; (v) a technology cannot be deployed in a country before the year ITU first records positive coverage for it, after which an exponential maturity ramp governs uptake; (vi) the footprint radius depends on technology and an urban/rural setting flag rather than on terrain-resolved radio propagation; and (vii) backhaul cost is the lesser of distance-scaled fibre to the nearest existing tower and a flat microwave link. These assumptions make Track~B interpretable and data-efficient where labels are sparse, at the price of bias where real deployment departs from profit-greedy rollout (e.g.\ universal-service obligations or state-led builds).

\subsection*{Track C: pix2pix spatial deep learning}\label{sec:trackC}

We frame coverage prediction as conditional image translation and train a U-Net generator in the pix2pix manner \cite{isola2017,ronneberger2015}, one model per technology. The pix2pix PatchGAN discriminator is implemented in the codebase, but the released models are trained with the adversarial term disabled, so the generator is supervised by its reconstruction losses alone. The generator ingests a 256$\times$256\,km patch of the per-country covariate stack: a nine-channel tensor of per-country-normalised population, road length, built-up surface, terrain ruggedness, elevation, SEDAC deprivation and nighttime lights, plus a $\sin/\cos$ encoding of the prediction year; and outputs the per-pixel coverage probability for that patch and year. The channel set is smaller than Track~A's 15 features: it keeps only the spatially-structured rasters and omits the country-level scalars (the two World Bank telecom indicators), which are constant within a country and therefore carry no spatial signal for a convolutional model, whereas the tabular learner exploits their variation across countries. Training minimises a weighted sum of binary cross-entropy (weight 1.0) and a boundary-aware Dice term (0.5), using the Adam optimiser at learning rate $10^{-4}$ with on-the-fly augmentation (horizontal/vertical flips at $p{=}0.5$ and $\pm 15^\circ$ rotation). Models train for 30 epochs at batch size 16. Inference uses Monte-Carlo dropout with $T{=}30$ stochastic forward passes \cite{gal2016} for per-pixel uncertainty (full settings in Table~\ref{tab:hyperparams}). Edge patches are zero-padded to a uniform shape and masked out in the loss.

\subsection*{Technology availability gate}\label{sec:gate}

A model fitted on presence/absence labels cannot know that a technology did not exist in a country in a given year. Left unconstrained it extrapolates coverage backwards from features that resemble later covered areas, so every track is gated on the year each technology first became commercially available in each country.

An earlier construction derived that year from the ITU population-coverage panel, taking the first year a country reports at least $1\%$ population coverage. That is a diffusion milestone rather than a launch, and it runs systematically late. Compared with independently sourced commercial launch years it is late by a median of 4 years for 2G (late in 159 of 183 comparable countries), 2 years for 3G (150 of 193), and unbiased for 4G (median 0, exact in 113 of 200). The bias decays by generation as that construction predicts: reaching $1\%$ of a population takes years of build-out from a standing start for 2G, but is close to immediate for 4G overlaid on existing sites. The ITU panel also begins in 2000, so no ITU-derived gate can precede that year, although GSM launched commercially in 1991.

We use web-sourced commercial launch years wherever they could be established to a high evidential standard via two separate searches on operator records, national regulators, industry databases and contemporaneous trade press, each seeking the first year service was sold to the public, as distinct from trials, licence awards or spectrum auctions. Where the searches differed by a single year, we chose the later candidate.

427 of 642 (country, technology) cells are certified in this way (2G 110, 3G 137, 4G 180; 196 countries), each carrying at least one citable source. The remaining 215 fall back on the ITU-derived year, taking the first year a country reports at least $1\%$ population coverage or alternatively on a global technology floor, the year each technology first became commercially available anywhere (2G 1995, 3G 2002, 4G 2010). Certified years are applied \emph{before} the reliable-label condition, which lowers any gate to the earliest year a reliable label proves coverage existed. The condition is therefore the final safety net: no year for which we hold positive evidence can be forbidden, whichever source supplied the gate.

\subsection*{Track combination and uncertainty}\label{sec:bestestimate}

\begin{figure}[!ht]
\centering
\includegraphics[width=\textwidth]{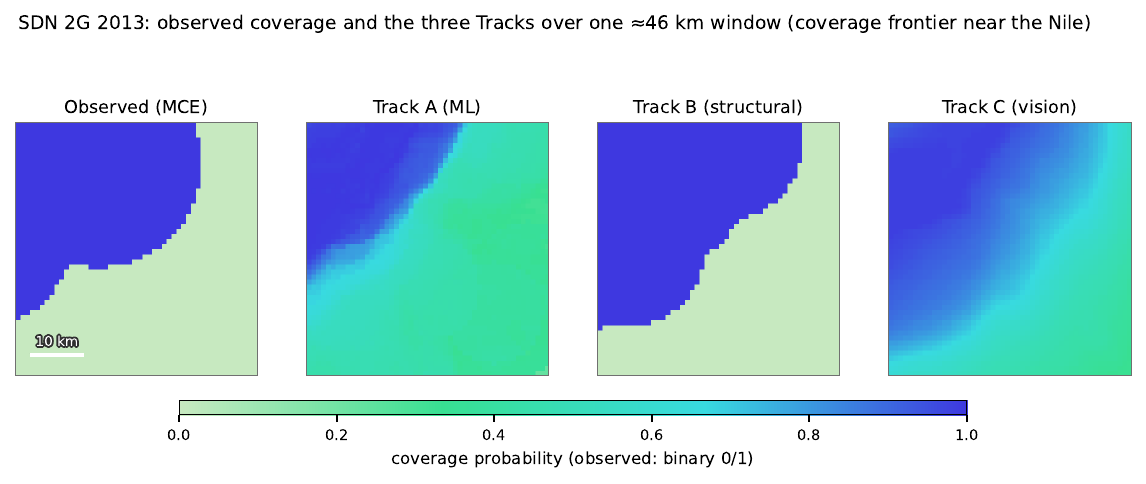}
\caption{The observed reliable map and the three modelling tracks over one
identical $\approx$46\,km window on a coverage frontier (Sudan~2G, 2013, along the Nile north of Khartoum). 2013 is a reliable observed year for Sudan~2G, so all four panels share the same (country, technology, year) and can be read pixel-for-pixel. \textbf{Observed (MCE)} is the binary ground-truth coverage map (covered vs.\ uncovered). \textbf{Track~A} yields a smooth continuous probability surface; \textbf{Track~B} places discrete antenna sites, so its coverage is a union of hard-edged footprints; \textbf{Track~C} produces a segmentation-style field with learned boundaries. Each track represents the same underlying coverage edge differently; the best-estimate layer is their per-pixel weighted mean (Section~\ref{sec:bestestimate}), so these complementary structures are combined rather than viewed separately. The 10\,km scale bar applies to all four.}
\label{fig:tracks}
\end{figure}

The three \emph{calibrated} per-track probabilities are combined per pixel into the best-estimate layer (Layer~4) as a weighted mean $\bar p=\sum_k w_k p_k$ ($k\in\{$A,B,C$\}$). For each (country, technology) we score each track's prediction against that group's reliable cells with the Brier score $L_k$. We choose the Brier score, i.e. the mean squared error against the observed coverage fraction, as the rule of choice, since, unlike log-loss, it is well behaved for Track~B's hard $0/1$ output. We set the weights to a tempered softmax of negative loss, $w_k \propto \exp(-L_k/\tau)$ with $\tau{=}0.05$, renormalised over the tracks present and given a small uniform floor so no track is dropped entirely. A track that is locally accurate, typically Track~A, therefore dominates the blend, while a locally unreliable one is down-weighted rather than discarded. Before combination, each track's per-pixel time series (and afterwards the combined mean itself) is made non-decreasing in year by a running-maximum transform; together with Track~A's monotone-increasing year constraint this encodes the assumption that deployed coverage is not torn down. The released layers therefore track \emph{maximum-achieved} coverage: genuine technology sunsets (2G/3G switch-offs, which began in some markets in the 2010s) are not represented, and per-technology values in late years should be read as ``coverage if the network were still operating'' (see Section \ref{sec:usage}). The three tracks attack the same prediction from different directions and produce qualitatively different coverage fields, which the weighted mean blends rather than shows in isolation; Figure~\ref{fig:tracks} compares them against the observed reliable map over a single window.

The released $90\%$ uncertainty interval is produced by conformal-style empirical residual calibration on the combined mean \cite{vovk2005,lei2018}: on the reliable label cells we compute signed residuals between $\bar p$ and the observed coverage fraction of the cell, take the finite-sample $5\%$/$95\%$ residual quantiles per stratum (technology $\times$ era $\times$ World-Bank region $\times$ income group, pooled up to coarser strata below a minimum cell count), and release $[\bar p + q^{\mathrm{lo}},\ \bar p + q^{\mathrm{hi}}]$ clipped to $[0,1]$. Because the same reliable cells also train the tracks and fit the combination weights, this is not a held-out split: the guarantee is empirical-in-distribution rather than the strict split-conformal one, and the leave-country-out check in Technical Validation (Table~\ref{tab:interval_coverage}) is the out-of-sample evidence for the released bands. No per-track variance is propagated. For label-free countries and projected years the interval is an extrapolation, as discussed in Section \ref{sec:usage}. The per-track layers ship as point estimates; their reliability is summarised by the validation tables rather than per-pixel uncertainty bands. For users who prefer a single model, \texttt{index.csv} records, per (country, technology), the track with the lowest held-out Brier. The released rasters are model output everywhere: observed MCE labels are never substituted into the product (only the reliability metadata is distributed, see Section~\ref{sec:datarecords}).

\subsection*{Code availability}\label{sec:code}

The full pipeline (acquisition, feature engineering, three tracks, validation harness, dataset packager) is available as a Git repository under BSD-3-Clause at \url{https://github.com/Societal-Computing/coverage_maps} and will be archived upon acceptance. The code accompanies the dataset for reproduction and audit. It is installable into the provided environment (\texttt{pip install -e .}) but is not distributed as a maintained package. Reproducing the dataset from raw inputs is a single \texttt{python -m scripts.run\_global} invocation, given the inputs and licences are obtained by the user.

\section*{Data Records}\label{sec:datarecords}

The product is deposited on Zenodo at \url{https://doi.org/10.5281/zenodo.21594337} under CC-BY-4.0. The rasters are distributed as 22 ZIP archives, one per UN M49 subregion, so that a user fetches only the regions they need; \texttt{regions.csv} maps each country to its archive. Within an archive the tree is organised by layer and technology:

{\small\begin{verbatim}
<subregion>.zip     22 archives, e.g. western_europe.zip
  combined/{tech}/{ISO3}_..._v1.tif  best estimate: 3-band COG
                        (mean, conformal 5% lower, conformal 95% upper)
  ml/{tech}/{ISO3}_..._v1.tif        Track A mean prob.  (1-band COG)
  structural/{tech}/{ISO3}_...       Track B mean prob.  (1-band COG)
  vision/{tech}/{ISO3}_...           Track C mean prob.  (1-band COG)
regions.csv           country -> subregion archive lookup
manifest.csv          per archive: size, file count, members, SHA-256
index.csv             per-(country,tech) Brier, BMA weight, best track
reliable_coverages.csv      label metadata; 2,488 rows,
                            2,409 distinct (country,tech,year) maps
reliable_legacy_report.csv  per-check PASS/FAIL only (no values)
launch_year_certification.csv  certified commercial launch year per cell
launch_year_provenance.csv     its source-level evidence
population_total_correction.csv  countries whose 2021-2030 national
                                 totals were rescaled
supplementary.zip     sensitivity, config_snapshot
metadata.yaml  CITATION.cff  README.md
\end{verbatim}}

The primary \texttt{combined/} file is a Cloud-Optimised GeoTIFF (COG) \cite{cog_spec}: a GeoTIFF internally tiled and overview-pyramided for efficient partial reads over the network holding the three-track combined estimate and its conformal prediction interval, in three float32 bands (EPSG:4326):

\begin{itemize}
\item \textbf{band 1: combined mean} coverage probability,
      $\bar p_{c,t,y}=\sum_{k\in\{A,B,C\}} w_k\,p_k$, a weighted average of
      the three \emph{calibrated} per-track probabilities. Weights $w_k$
      are per-(country, technology), a tempered softmax of each track's
      negative Brier loss on the reliable labels (Methods,
      Section~\ref{sec:bestestimate});
\item \textbf{bands 2--3: split-conformal lower / upper bounds},
      $[\bar p + q^{\mathrm{lo}},\ \bar p + q^{\mathrm{hi}}]$ clipped
      to $[0,1]$, where $q^{\mathrm{lo}}, q^{\mathrm{hi}}$ are
      per-stratum signed residual quantiles at the $90\%$ level,
      calibrated on held-out reliable pixels (Methods,
      Section~\ref{sec:bestestimate}).
\end{itemize}

\begin{figure}[p]
\centering
\includegraphics[width=\textwidth,height=0.88\textheight,keepaspectratio]{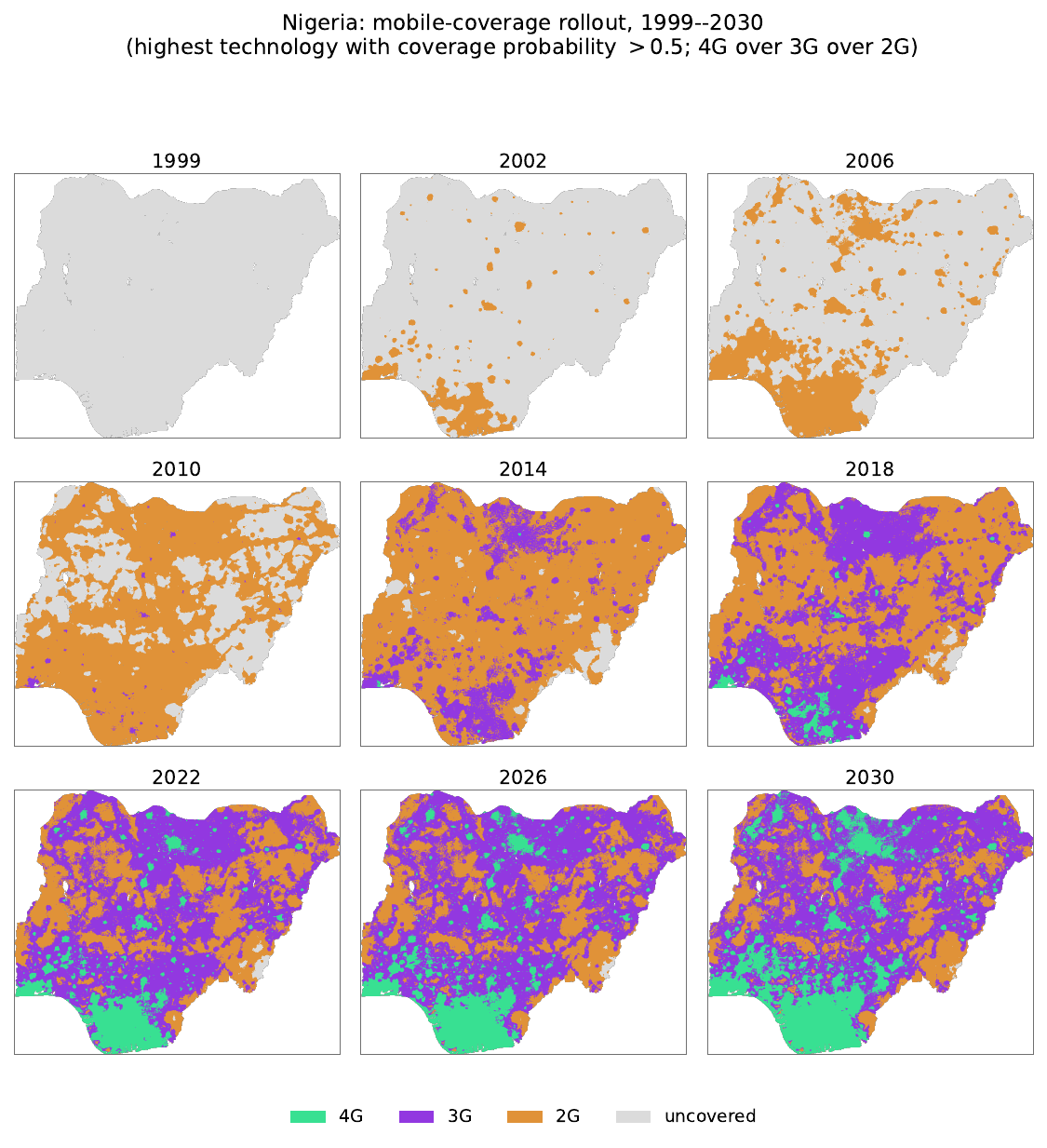}
\caption{Mobile-coverage rollout in Nigeria, 1999--2030, from the released
best-estimate layers (one panel per year). A pixel is coloured by the highest technology whose combined coverage probability exceeds $0.5$ (\textcolor{cGreen}{4G} over \textcolor{cPurple}{3G} over \textcolor{cBrown}{2G}) and left grey where no technology reaches $0.5$. The sequence makes the core value of the dataset visible: a consistent 1\,km, multi-technology record of where and when coverage arrived, extended over the 2026/2030 projection horizon. 2G appears first and reaches furthest toward the rural frontier; 3G and then 4G fill in from population centres outward, with 4G covering most of the populated country by the late 2020s while 2G persists where the newer technologies have not yet reached.}
\label{fig:rollout}
\end{figure}

Figure~\ref{fig:rollout} showcases the dataset's central contribution: a temporally-consistent, multi-technology coverage record through the 1999--2030 rollout of 2G, 3G and 4G in Nigeria.

The per-track \texttt{ml/} (Track~A), \texttt{structural/} (Track~B) and \texttt{vision/} (Track~C) directories provide each model's mean probability as a single-band COG, for users who prefer one track over the average. The per-track layers are point estimates: calibrated uncertainty is provided only on the combined product, and per-track reliability is reported in the validation tables (Technical Validation). The label-set metadata ships in \texttt{reliable\_coverages.csv}; \texttt{index.csv} records, per (country, technology), each track's held-out Brier score, its BMA weight, and the single recommended track, for users who prefer one model over the combined average (Section~\ref{sec:bestestimate}).

\emph{Country coverage.} The dataset spans all 214 reference-grid countries. Of these, 207 carry reliable-triple labels for at least one technology; the remaining 7 (Comoros, Djibouti, North Korea, Palestine, Saint Helena, South Sudan and Kosovo) have no reliable triple and are covered by the modelling tracks alone. Track~B has parameters for all 214: 191 fitted directly by CMA-ES and 23 pooled from income-group donors where a country's own fit found no signal. Reliability is moreover uneven \emph{within} countries: a country labelled reliable for one technology (most often 2G) frequently lacks a reliable triple for another (typically 4G), reflecting the era tilt of the underlying archive (Section~\ref{sec:reliability}). Such unlabelled (country, year, technology) cells rest on the modelling tracks alone, exactly as the seven fully-unlabelled countries do, so reliable-label coverage should be read per technology rather than per country.

\texttt{index.csv} summarises, per (country, technology), each track's held-out Brier, its BMA weight and the recommended single track. \texttt{reliable\_coverages.csv} documents the supervised label set for audit: 1{,}859 country-year rows encoding the 2{,}488 reliable triples (publication years, reliability flags and \texttt{source}), thereof1{,}058 modern-MCE and 1{,}430 legacy-archive, distinguished by the \texttt{source} column. \texttt{config\_snapshot/} embeds the yaml configs (grid, reliability, covariates, tracks) that produced the release.

\section*{Technical Validation}\label{sec:validation}

\subsection*{Spatial 5-fold cross-validation}\label{sec:internalcv}

Track~A's primary generalisation estimate is a 5-fold spatial cross- validation across the 207 countries with usable reliable-triple data for at least one technology (see Section~\ref{sec:reliability}). The per-technology training sets span 204 (2G), 142 (3G) and 97 (4G) countries, reflecting the progressively sparser reliable 3G/4G coverage record. Within each technology, countries are assigned to five disjoint round-robin folds. For each (technology, fold) we train a LightGBM booster on the remaining four folds' countries at the production configuration (a fixed 2\,000 boosting rounds, a 1~M-pixel-per-country deterministic uniform subsample and $1/\sqrt{n_c}$ per-country sample weights, see Methods, Section~\ref{sec:trackA}) and evaluate on the held-out fold. For computational efficiency the 2G folds' training rows (by far the largest label set, Section~\ref{sec:reliability}) are uniformly subsampled to 100~M before fitting; the 3G and 4G folds use all available rows. To measure calibration out-of-distribution, the per-fold isotonic calibrator is fit on a 5\,\% slice held out of the \emph{training} fold and every metric is then scored on the \emph{entire} held-out fold. This is stricter than the more common practice of fitting the calibrator on part of the evaluation fold: the calibrator never sees the held-out countries' distribution, so the reported expected calibration error reflects genuine transfer to unseen countries.

As a floor for these numbers, a single-covariate baseline that ranks pixels by population density alone scores AUC 0.673 / 0.679 / 0.710 (2G/3G/4G) on the same reliable labels, and built-up surface alone scores 0.670 / 0.676 / 0.677. Track~A sits clearly above the population floor (see Table \ref{tab:trackA_cv}). The margin is widest for 2G and narrowest for 4G: 4G coverage is the most concentrated in dense settlement, so population alone predicts it best, and the multi-covariate booster has correspondingly less left to add. Even at its narrowest the margin is substantial, and it confirms that the boosters capture coverage structure beyond where people simply live.

\begin{table}[!ht]
\centering
\small
\resizebox{\textwidth}{!}{\begin{tabular}{lcccccc}
\toprule
\textbf{Tech} & \textbf{Folds} & \textbf{AUC} & \textbf{Median AUC} &
\textbf{Brier} & \textbf{ECE} & \textbf{Dice} \\ \midrule
2G & 5 & 0.917 $\pm$ 0.026 & 0.925 & 0.117 $\pm$ 0.027 & 0.051 $\pm$ 0.037 & 0.769 $\pm$ 0.046 \\
3G & 5 & 0.896 $\pm$ 0.009 & 0.899 & 0.127 $\pm$ 0.016 & 0.063 $\pm$ 0.049 & 0.715 $\pm$ 0.049 \\
4G & 5 & 0.885 $\pm$ 0.028 & 0.875 & 0.139 $\pm$ 0.031 & 0.090 $\pm$ 0.061 & 0.596 $\pm$ 0.143 \\
\bottomrule
\end{tabular}}
\caption{Track~A 5-fold spatial cross-validation with out-of-distribution
calibration (the per-fold isotonic calibrator is fit on a held-out training-fold slice, never on the evaluation fold; every metric is scored on the full held-out fold). AUC ranges 0.89--0.92 across technologies, with 2G highest (0.92) and 3G/4G behind at 0.90 and 0.89; the country-block fold structure exposes real heterogeneity in spatial generalisation (per-fold AUC std 0.01--0.03). Dice is lower for 3G and 4G (0.72 and 0.60) than for 2G (0.77) because their per-fold positive base rates are lower (22--47\,\% for 4G), so the 0.5-threshold binary decision is strictly harder than the underlying probabilistic ranking captures; the AUC is the more representative metric in that regime. Out-of-distribution expected calibration error (ECE) has median 0.040 / 0.041 / 0.104 (2G/3G/4G); the tabulated means (0.051 / 0.063 / 0.090) are inflated by a single high-positive-rate fold per technology (e.g.\ 4G fold~5: 47\,\% positive, ECE 0.28).}
\label{tab:trackA_cv}
\end{table}

\subsection*{Temporal hold-out validation}\label{sec:temporalcv}

The spatial cross-validation above tests generalisation across \emph{countries} but not across \emph{time}, yet the product's central deliverable is a 1999--2030 annual series. We therefore run a temporal hold-out: for each technology a LightGBM booster is trained on all reliable rows with year $\le 2015$ and evaluated on held-out rows with year $\ge 2016$ (training rows again subsampled to at most $100$ M).

\begin{table}[!ht]
\centering
\small
\begin{tabular}{lccccc}
\toprule
\textbf{Tech} & \textbf{train $\le$2015} & \textbf{test $\ge$2016} &
\textbf{AUC} & \textbf{Brier} & \textbf{pos.\ rate} \\ \midrule
2G & 100.0\,M & 5.0\,M & 0.944 & 0.093 & 61.9\,\% \\
3G & 100.0\,M & 5.0\,M & 0.926 & 0.107 & 43.5\,\% \\
4G & 14.7\,M & 5.0\,M & 0.891 & 0.126 & 34.9\,\% \\
\bottomrule
\end{tabular}
\caption{Temporal hold-out: train on years $\le 2015$, evaluate on
$\ge 2016$. Forward-in-time AUC (0.891--0.944) matches or exceeds the spatial cross-validation (Table~\ref{tab:trackA_cv}). The two tests are complementary: this one isolates \emph{year}-extrapolation for countries seen during training, whereas the spatial CV isolates generalisation to \emph{unseen countries} so the higher temporal AUC is expected (the held-out years belong to familiar countries) and should not be read as Track~A being better in time than in space. The result supports use of the dataset across its annual steps through 2024; the projected 2025--2030 tail remains a covariate-driven extrapolation beyond any label (see Section \ref{sec:usage}).}
\label{tab:temporal}
\end{table}

\subsection*{External validation against ITU country aggregates}

For each (country, year, technology) we aggregate predicted per-pixel coverage to a population-weighted country share and compare it against the ITU DataHub population-coverage indicators~\cite{itu_datahub} (\emph{i}~$\in$ \{100093, 100094, 100095\} for at-least-2G/3G/4G).

\begin{table}[!ht]
\centering
\small
\resizebox{\textwidth}{!}{\begin{tabular}{lcccccc}
\toprule
\textbf{Tech} & \textbf{n} & \textbf{Pearson r} & \textbf{Spearman} $\boldsymbol{\rho}$ & \textbf{MAE} &
\textbf{mean predicted} & \textbf{mean ITU} \\ \midrule
\multicolumn{7}{l}{\textit{Combined best estimate (BMA)}}\\
2G & 1\,798 & 0.618 & 0.624 & 0.051 & 0.927 & 0.950 \\
3G & 1\,841 & \textbf{0.761} & \textbf{0.784} & 0.127 & 0.749 & 0.782 \\
4G & 1\,427 & \textbf{0.831} & \textbf{0.840} & 0.154 & 0.511 & 0.590 \\
\midrule
\multicolumn{7}{l}{\textit{Track A (LightGBM, calibrated)}}\\
2G & 1\,798 & \textbf{0.644} & \textbf{0.661} & 0.054 & 0.922 & 0.950 \\
3G & 1\,841 & 0.734 & 0.769 & 0.145 & 0.723 & 0.782 \\
4G & 1\,427 & 0.816 & 0.839 & 0.162 & 0.497 & 0.590 \\
\midrule
\multicolumn{7}{l}{\textit{Track B (structural simulator)}}\\
2G & 1\,798 & 0.371 & 0.526 & 0.065 & 0.936 & 0.950 \\
3G & 1\,841 & 0.700 & 0.692 & 0.129 & 0.786 & 0.782 \\
4G & 1\,427 & 0.799 & 0.802 & 0.147 & 0.584 & 0.590 \\
\midrule
\multicolumn{7}{l}{\textit{Track C (pix2pix DL, MC-dropout mean)}}\\
2G & 1\,798 & 0.433 & 0.489 & 0.056 & 0.937 & 0.950 \\
3G & 1\,841 & 0.687 & 0.693 & 0.139 & 0.818 & 0.782 \\
4G & 1\,427 & 0.730 & 0.717 & 0.190 & 0.514 & 0.590 \\
\bottomrule
\end{tabular}}
\caption{External validation of the released combined layer and the three
per-track layers against ITU. Bold marks the best Pearson and Spearman value
for each technology across the four layers. For every layer we aggregate the released per-pixel COG to a population-weighted country-year coverage share and compare it with the ITU DataHub indicators 100093/4/5 (at-least-2G/3G/4G population coverage); all layers therefore use an identical method, sample and window ($\sim$1{,}400--1{,}840 country-years per technology over 2010--2020, all countries with both a COG and an ITU value). Note a definitional asymmetry: the ITU indicators are \emph{at-least-$X$} population shares (covered by $X$ \emph{or any newer} technology), whereas our layers are technology-specific, so wherever a newer technology's footprint exceeds the older one's, including every post-2G-sunset country-year, ITU reads structurally higher than an accurate tech-specific 2G/3G prediction; the 2G/3G correlations below are therefore conservative. The combined layer attains Pearson $r$ 0.62--0.83, Spearman $\rho$ 0.62--0.84 and MAE 0.05--0.15, with mean predicted within 2--8\,pp of mean ITU, and it is the strongest of the four on 3G and 4G. On 2G it is not: Track~A alone reaches $r$ 0.644 there, above the combined 0.618, and Track~A is also the strongest single track on 3G and 4G ($r$ 0.734 and 0.816) before combination. 2G is near-saturated, so the per-track loss spread that sets the BMA weights compresses and the two weaker auxiliary tracks retain more weight than their 2G accuracy warrants. Track~B (the techno-economic simulator) runs $r$ 0.37 / 0.70 / 0.80 and Track~C (vision) 0.43 / 0.69 / 0.73, both closing the gap as the technology gets newer. The high-coverage 2G/3G shares have little cross-country variance, so their Pearson values are attenuated by restriction of range and the rank ($\rho$) and absolute (MAE) agreement are more informative there. As an independent patch-level check, the released Track~C models attain held-out Dice $0.72/0.61/0.42$ (2G/3G/4G) on their per-technology validation countries ($31/21/15$); the local segmentation is thus spatially coherent even though the \emph{country-aggregate} 2G shares it produces are noticeably noisier, which is precisely why the BMA leaves Track~C down-weighted on 2G.}
\label{tab:external_validation}
\end{table}

\paragraph{Comparability with ITU.}
For this comparison our per-pixel coverage probabilities are aggregated to country-year shares as population-weighted means over the WorldPop population of each cell, so that both quantities are population-weighted percentages of inhabitants ``within range of a signal''~\cite{itu_facts_figures_2024} and the \emph{unit} is identical (see Table~\ref{tab:external_validation}). The residual per-(country,\,year) differences partly reflect a genuine definitional gap between our MCE-grounded target and the ITU indicator, not only model error: what differs is the operational definition of ``covered'':
\begin{itemize}
\item Our model is trained against the Mobile Coverage Explorer
(MCE) operator-submitted service-area
polygons~\cite{collins_bartholomew_mce}. MCE has nominal signal-
strength thresholds (2G/3G: $\geq -100$\,dBm strong / $\geq
-92$\,dBm variable; 4G: $\geq -120$\,dBm), though the documentation
acknowledges that data from many operators ``do not include signal
strength information or do not follow [these]
guidelines''~\cite{collins_bartholomew_mce}. Pixels are flagged
covered only where a submitting operator has explicitly declared a
service area.
\item The ITU indicator specifies no signal threshold at all
(``access to a mobile cellular
signal''~\cite{itu_facts_figures_2024}). It draws on regulator and
operator self-reports, and in many cases the data ``refer only to
the operator with the largest coverage''~\cite{itu_facts_figures_2024}.
Missing values are imputed from a five-year trend model in IBM
SPSS, and the urban-versus-rural split is derived by assuming
urban coverage is essentially complete.
\end{itemize}
Where ITU relies on imputation or single-operator reporting it tends to read higher than the MCE-grounded labels our model learns from; this explains part of the residual spread and the larger 3G/4G MAE. Because the aggregate offset is small, both the \emph{rank} correlations (trend agreement) and the absolute differences (close in aggregate) in Table~\ref{tab:external_validation} are informative, and we report both for transparency.

\subsection*{Effect of the technology availability gate}

The launch years were sourced from deployment records and never fitted to ITU, so their effect on agreement with ITU is an independent check on the gate itself. Recomputing Track~A's external validation with and without the gate improves agreement on every technology and on both metrics (Table~\ref{tab:gate_effect}). All six movements favour the gated series, which indicates that the pre-launch coverage the gate removes is spurious.

\begin{table}[!ht]
\centering
\small
\begin{tabular}{lcccc}
\toprule
 & \multicolumn{2}{c}{\textbf{Pearson $r$}} & \multicolumn{2}{c}{\textbf{MAE}} \\
\cmidrule(lr){2-3}\cmidrule(lr){4-5}
\textbf{Tech} & ungated & gated & ungated & gated \\ \midrule
2G & 0.800 & 0.806 & 0.0627 & 0.0621 \\
3G & 0.692 & 0.700 & 0.1466 & 0.1457 \\
4G & 0.783 & 0.792 & 0.1668 & 0.1639 \\
\bottomrule
\end{tabular}
\caption{Track~A agreement with ITU country aggregates before and after the
technology availability gate (Section~\ref{sec:gate}). Higher $r$ and lower
MAE indicate closer agreement. Only the ungated-to-gated difference within this table is meaningful.}
\label{tab:gate_effect}
\end{table}

\subsection*{Uncertainty interval coverage}

We validate the released $90\%$ split-conformal interval (Section~\ref{sec:bestestimate}) on the held-out reliable pixels, scoring the band $[\bar p+q^{\mathrm{lo}},\,\bar p+q^{\mathrm{hi}}]$ against the observed coverage \emph{fraction} of each $1$\,km cell. The calibration set comprises $15.2$\,M reliable cells drawn from $2{,}378$ reliable (country, technology, year) triples across the $207$ countries whose reliable cells could be sampled from the released grid. Two coverage estimates are reported (Table~\ref{tab:interval_coverage}): \emph{in-sample}, where the per-stratum residual quantiles are evaluated on their own calibration pixels (nominal by construction: $0.90$ for every technology); and \emph{spatially-blocked}, a $5$-fold leave-country-out approach in which the quantiles are refit on training-fold countries and the interval is scored on the held-out countries' pixels, which isolates generalisation to unseen countries.

Blocked coverage is $0.899$ overall and stays within $0.897$--$0.904$ across every technology$\times$era stratum, so the interval keeps its nominal level even on countries absent from the calibration fit. The mean band width quantifies how informative the interval is: it is narrow where the model is confident (e.g.\ European 2G, width $<0.10$) and wide where the per-pixel coverage fraction is genuinely hard to pin down from covariates alone (3G/4G and sparsely-labelled regions), giving a blocked-mean width of $0.63$ (2G), $0.64$ (3G) and $0.56$ (4G). The bands are wide because they are well-calibrated, not optimistic: reaching $90\%$ coverage of the continuous fraction requires them to absorb the model's residual mis-calibration, which is larger for the newer technologies. Country- and region-aggregated estimates (the typical use of the dataset) carry substantially smaller uncertainty than these per-pixel bands.

\begin{table}[!ht]
\centering
\small
\begin{tabular}{llrcc}
\toprule
\textbf{Tech} & \textbf{Era} & \textbf{calib.\ cells} &
\textbf{blocked coverage} & \textbf{mean band width} \\ \midrule
2G & pre-2010  &  8{,}235{,}398 & 0.898 & 0.620 \\
2G & post-2010 &  2{,}605{,}308 & 0.897 & 0.651 \\
3G & pre-2010  &    753{,}359 & 0.903 & 0.736 \\
3G & post-2010 &  2{,}368{,}463 & 0.899 & 0.615 \\
4G & post-2010 &  1{,}237{,}154 & 0.904 & 0.555 \\ \midrule
\multicolumn{3}{l}{\textit{overall (15.2\,M cells)}} & 0.899 & 0.625 \\
\bottomrule
\end{tabular}
\caption{Empirical coverage of the released $90\%$ split-conformal
interval on held-out reliable pixels, under a $5$-fold leave-country-out (spatially-blocked) approach. In-sample per-stratum coverage is $0.90$ by construction for every technology; the blocked figures shown here test generalisation to unseen countries and remain within $\pm0.6$\,pp of nominal. 4G has no pre-2010 reliable labels (4G post-dates the legacy archive). Band width is the mean of $(\mathrm{upper}-\mathrm{lower})$ over scored cells.}
\label{tab:interval_coverage}
\end{table}

\subsection*{Sensitivity analyses}

We assess how much of the product depends on individual design choices at the stage where those choices act, i.e. the label filters, by re-running the filters under perturbed settings and reporting the resulting change in label volume. Two families of perturbation are examined.

\emph{Modern reliability filter.} (i) Widening the operator-recency window from 2 years to 5 and 10 years; (ii) lowering the major-operator tower-count threshold from 75\,\% to 50\,\%; and (iii) dropping the curated additional-majors override entirely. At the release settings the modern filter admits 1{,}058 reliable (country, tech, year) triples across 168 countries. Widening the recency window is the only perturbation that moves label volume materially (to 1{,}155 triples ($+97$) at 5 years and 1{,}200 ($+142$) at 10 years) as it readmits country-years whose most recent qualifying operator submission predates the 2-year window. Lowering the major-operator threshold from 75\,\% to 50\,\% and dropping the curated overrides both leave the admitted set unchanged (1{,}058 triples): the operator-majority step is not the binding constraint on which MCE country-years qualify. The released $2$-year window is thus the conservative choice on the only sensitive axis.

\emph{Legacy substitute filter.} Because the pre-2010 archive is admitted through three independent consistency checks (Section~\ref{sec:reliability}), we vary each threshold in turn (ITU agreement $\pm 25$\,pp $\rightarrow \pm 15 / \pm 35$\,pp, temporal monotonicity $0.95 \rightarrow 0.90 / 0.98$, and 2010-boundary overlap $0.80 \rightarrow 0.70 / 0.90$) re-deriving acceptance from the filter's per-check records. Acceptance moves little: from 73.4\,\% (1{,}430 triples) at the release settings to between 70.4\,\% (tightest ITU tolerance, $-60$ triples) and 76.2\,\% (loosest, $+53$); the monotonicity and boundary perturbations shift acceptance by between $-49$ / $+38$ and $-16$ / $+13$ triples respectively (Table~\ref{tab:legacy_sens}). The 1999--2009 extension is thus not an artefact of the substitute filter's calibration. The computed per-setting deltas (Table~\ref{tab:legacy_sens}) ship in the deposit's \texttt{sensitivity/} directory.

\begin{table}[!ht]
\centering
\footnotesize \setlength{\tabcolsep}{4pt}
\begin{tabular}{lcccc}
\toprule
\textbf{Variant} & \textbf{Accepted} & \textbf{Rate} &
\textbf{$\Delta$ vs release} & \textbf{(2G / 3G)} \\ \midrule
Release ($\pm$25\,pp / 0.95 / 0.80) & 1{,}430 & 73.4\,\% & --   & 1{,}325 / 105 \\
ITU tolerance $\pm$15\,pp           & 1{,}370 & 70.4\,\% & $-60$ & 1{,}276 / 94  \\
ITU tolerance $\pm$35\,pp           & 1{,}483 & 76.2\,\% & $+53$ & 1{,}369 / 114 \\
Monotonicity floor 0.90             & 1{,}468 & 75.4\,\% & $+38$ & 1{,}362 / 106 \\
Monotonicity floor 0.98             & 1{,}381 & 70.9\,\% & $-49$ & 1{,}276 / 105 \\
Boundary overlap 0.70               & 1{,}443 & 74.1\,\% & $+13$ & 1{,}336 / 107 \\
Boundary overlap 0.90               & 1{,}414 & 72.6\,\% & $-16$ & 1{,}311 / 103 \\
\bottomrule
\end{tabular}
\caption{Sensitivity of legacy-triple acceptance to the substitute
filter's thresholds, re-derived from the filter's per-check records (1{,}947 candidates). Each row perturbs one threshold, holding the others at release settings.}
\label{tab:legacy_sens}
\end{table}

\emph{Training-set subsampling.} For tractability on commodity hardware, the reported cross-validation (Table~\ref{tab:trackA_cv}) trains each fold on a uniform $100$ M-row subsample of the per-country-capped spill (the full 2G fold is $\approx$0.5\,billion rows). To confirm this subsample does not distort discrimination, we recomputed one 2G fold at its full $\approx$$5.4\times10^8$ rows: the held-out-fold AUC moves from $0.928$ ($10^8$ cap) to $0.925$ (full fold), a shift of $0.003$, an eighth of the per-fold AUC standard deviation in Table~\ref{tab:trackA_cv}, and within run-to-run subsample noise. The $10^8$ cap is therefore immaterial to the booster's ranking performance. The release uses the primary configuration throughout.

\begin{figure}[p]
\centering
\includegraphics[width=\textwidth,height=0.74\textheight,keepaspectratio]{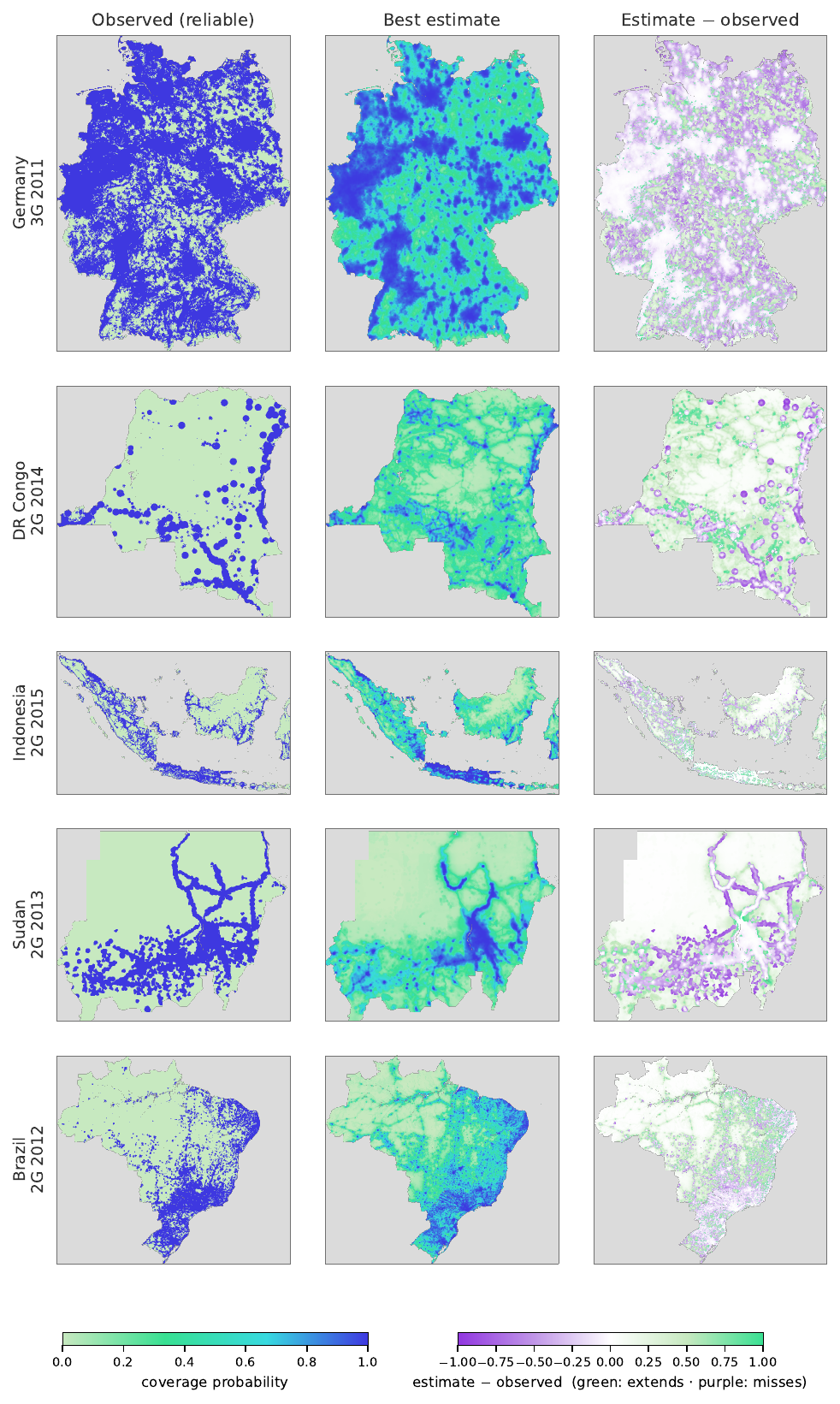}
\caption{Case-study coverage mosaics for five countries (rows), each at a
reliable (technology, year) snapshot annotated at left: Germany~3G~2011, DR~Congo~2G~2014, Indonesia~2G~2015, Sudan~2G~2013 and Brazil~2G~2012. Columns: the observed reliable map, the best-estimate combined prediction, and their difference (estimate~$-$~observed). Out-of-footprint cells are left grey. Coverage probability uses a sequential light-green-to-dark-blue scale (the observed map is shown as its binary 0/1 on the same scale); the difference uses a diverging scale centred on white, \textcolor{cGreen}{green} where the model \emph{extends} coverage beyond the observed label and \textcolor{cPurple}{purple} where it \emph{misses} observed coverage. Indonesia is cropped to its populated western core (Sumatra--Java--Borneo--Sulawesi) for legibility. The cases span dense near-complete coverage (Germany), sparse low-coverage interiors (DR~Congo, Sudan), an archipelago (Indonesia) and a large heterogeneous deployment (Brazil), showing where the modelled layers extend coverage beyond the observed labels (broad green areas in DR~Congo and Brazil) and where they fall short of dense observed coverage (purple along Sudan's Nile corridor).}
\label{fig:case_studies}
\end{figure}

Figure \ref{fig:case_studies} shows observed, best-estimate, and difference mosaics for five countries spanning easy and hard regimes: Germany (mature, dense reliable labels; 3G~2011), DR~Congo (sparse reliable labels, low coverage; 2G~2014), Indonesia (archipelago, multi-operator; 2G~2015, populated western core shown), Sudan (sparse infrastructure; 2G~2013) and Brazil (large, heterogeneous deployment; 2G~2012). Each country is shown at a reliable (technology, year) snapshot where coverage is substantial but incomplete, which makes the difference panel informative as near-fully-covered snapshots would leave it almost empty.

\section*{Usage Notes}\label{sec:usage}

The dataset is designed for three families of use. First, \emph{mapping the
digital divide}: coverage probabilities can be aggregated to any sub-national
geography (districts, catchments, survey clusters) to quantify who remained
unconnected, where, and for how long. Second, \emph{linkage with georeferenced
micro-data}: joined to household surveys (DHS, MICS, LSMS), census micro-data
or mobile-money records, the per-technology \emph{arrival timing} enables
event-study designs on the societal effects of connectivity; for causal work
we recommend restricting identification to the observed label subset
distributed in \texttt{reliable\_coverages.csv} (see the circularity note
below). Third, \emph{operational planning}: humanitarian access mapping,
digital-intervention targeting, and universal-service costing. Users should
read the era vocabulary used throughout: \textbf{label-anchored}
(1999--2020, ground truth exists for a screened subset of country-years),
\textbf{modelled} (2021--2024, covariates observed but operator maps sparse)
and \textbf{projected} (2025--2030, covariates themselves projected or held
forward; treat as scenario, not measurement).

\begin{itemize}
\item \textbf{Coverage is non-decreasing by construction.} Every released
      layer is constrained to be non-decreasing in year. The product therefore records \emph{maximum-achieved} coverage per technology: 2G/3G network sunsets are \emph{not} represented, and late-year 2G/3G values in
      such markets should be read as historical-maximum footprints, not
      operating networks. Analyses of technology retirement need an external
      sunset source.
\item \textbf{Observed-grounded vs projected years.} Reliable MCE
      labels exist only through 2020. Predictions for 2021--2024 are
      covariate-grounded but label-free, and 2025--2030 are a forward
      \emph{projection}: nighttime lights are held flat at their 2024
      value and population holds the 2020 allocation fixed, so no
      post-2024 quantity is observation-constrained. Treat the
      2025--2030 slices as a model extrapolation conditioned on
      projected covariates, not as measured coverage. No year beyond 2020
      is label-supported in \texttt{reliable\_coverages.csv}. Analyses of \emph{trends} should not
      cross the 2020/2024 observation horizon without a sensitivity
      check on the projected tail.
\item \textbf{Launch-year provenance is uneven and is published per
      cell.} 215 of 642 (country, technology) cells could not be
      certified against deployment records and are thus alternatively gated by ITU-derived year or alternatively the global
      technology floor instead. The launch-year sources ship with the dataset, so any
      individual gate can be checked rather than taken on trust.
\item \textbf{The gate is a single first year and cannot represent
      interrupted service.} North Korea is the clearest case: a GSM
      network operated from November 2002 but was shut down in 2004,
      with handsets effectively banned until 2008. A single first-year
      gate would wrongly permit continuous coverage across that gap, so
      DPRK 2G is left uncertified even though its launch year is well
      evidenced. ``Commercial launch'' is itself ambiguous where a
      network opened in the capital first, or where one source means GSM
      by ``2G'' and another means any digital cellular.
\item \textbf{Satellite direct-to-cell connectivity is out of scope.} The
      product maps \emph{terrestrial} cellular network coverage. Low-earth-orbit
      constellations with direct-to-cell service (e.g. Starlink/T-Mobile, AST
      SpaceMobile) began commercial operation in the mid-2020s and may, within
      years, provide basic connectivity independent of terrestrial towers. The
      projected 2025--2030 layers do not model this channel; users studying
      the near future should treat ``uncovered'' as ``uncovered by terrestrial
      networks.'' For the retrospective 1999--2024 core, the label-anchored
      period this dataset is primarily built for, the distinction is irrelevant.
\item \textbf{Population layers} span 1999--2030 from three regimes:
      years before 2000 reuse the earliest available WorldPop surface
      (year 2000); 2000--2020 are unconstrained UN-adjusted WorldPop;
      2021--2030 keep the 2020 allocation, rescaled so that each
      year matches its national total from WorldPop Global-2 R2024B
      (unconstrained, UN-adjusted). Only the national totals of that
      release are used; its gridded allocation is not. ``UN-adjusted''
      and ``constrained'' are independent WorldPop attributes: both
      series scale national totals to UN World Population Prospects,
      but the \emph{constrained} series allocates population only to
      pixels within mapped settlement footprints, whereas the
      \emph{unconstrained} series spreads it over all land pixels.
      The methodology break at the 2020/2021 boundary is therefore
      the constrained/unconstrained switch; users
      running difference-in-difference designs across this boundary
      should add a year-2021 fixed effect, and should treat the
      pre-2000 layer as held flat.
\item \textbf{Acute supply shocks are not modelled.} The model
      assumes coverage follows its structural covariates (population,
      built environment, roads, nighttime lights), so it cannot
      represent abrupt infrastructure loss from conflict or disaster
      such as in Sudan's recent war, where mobile infrastructure was
      destroyed faster than, and independently of, any covariate
      change. In such settings the estimates (and their intervals,
      which are calibrated on non-conflict data) should be read as
      upper bounds on actual coverage.
\item All summary statistics involving area (e.g.\ ``\% land
      covered'') should be computed using the \texttt{pixel\_area
      \_km2} column, not on raw pixel counts, because the EPSG:4326
      grid does not preserve area at high latitudes.
\item Track~A is the recommended single-track product for users
      without strong methodological preferences (smallest mean
      absolute calibration error). Track~B is preferred where
      interpretability of the coverage drivers matters (the
      calibrated $\alpha_{c,t}$ recover a per-country measure of
      tower-deployment friction). Track~C is preferred where local
      spatial morphology of coverage edges is the primary signal.
\item \textbf{Use one source consistently in panels.} For
      sub-national or panel analyses, use the
      combined-mean band (band~1 of \texttt{combined/}) throughout. Users with
      their own MCE access should not splice observed MCE maps (at
      reliable country-years) with model predictions (elsewhere):
      observed and modelled coverage differ in definition and error
      structure, so switching sources across years or countries
      introduces spurious jumps that masquerade as coverage change.
\item \textbf{International-border buffer under-count.} The
      \texttt{pop\_buffer\_10km} and \texttt{pop\_x\_arpu\_buffer
      \_10km} features used by Tracks~A and~B are computed by 2-D
      circular convolution of each country's WorldPop raster. Cross-
      border population is therefore \emph{not} counted: a pixel
      $\leq 10$\,km from an international land border sees only its
      own country's neighbours in the buffer, biasing the buffer
      value downward geometrically and is largest where two operators competed across a
      dense conurbation that straddles a national border (US--Mexico
      at El Paso--Ciudad Ju\'arez, EU internal borders post-Schengen,
      Hong Kong--Shenzhen). For most border strips, i.e. ocean coasts,
      desert frontiers, mountainous interior borders, the under-
      count is negligible. Users running border-region analyses
      should either (i) regenerate the buffer layer from a multi-
      country WorldPop mosaic clipped to the analysis-region $+$
      $10$\,km buffer, or (ii) restrict their analytical window to
      pixels at $\geq 10$\,km from any international border (a
      pre-computed border-distance helper is available in
      \texttt{coverage\_gaps.features.\allowbreak geometry.\allowbreak border\_distance}).
      Internal (sub-national) state and province borders carry
      \emph{no} corresponding artefact because all such neighbours
      are within the same country.
\item \textbf{Country imbalance in supervised tracks.} Pixel counts
      vary by $>\!4$ orders of magnitude between countries (Andorra
      $\sim\!10^3$, Russia $\sim\!10^7$). Track~A counteracts this with
      $1/\sqrt{n_c}$ per-row sample weighting and a $10^6$-pixel-per-
      country annual cap, so the global objective is not dominated by
      the giants; the cross-validation in Table~\ref{tab:trackA_cv} is
      computed under this weighting. Users producing global aggregates
      should nonetheless weight by \texttt{pixel\_area\_km2} rather
      than raw pixel counts.
\end{itemize}

\begin{appendices}

\section{Full technical specifications}\label{app:specs}

This appendix collects the complete technical settings referenced from Methods.

\paragraph{Operator-name matching.} Major operators are extracted
from the OpenCellID MNC table and matched to MCE \texttt{NAME} entries by lower-cased, iso2-scoped substring matching after applying a 218-entry brand-normalisation list covering known cross-database spelling variants (e.g.\ KT/olleh in Korea, MTN Cameroon $\to$ MTN, Vodafone India $\to$ Vi); research notes and source citations are embedded alongside each entry in the released configuration.

\begin{table}[!ht]
\centering
\small
\begin{tabular}{ll}
\toprule
\textbf{Country} & \textbf{Override operators (added as major)} \\ \midrule
Bangladesh   & Robi, Banglalink \\
Cameroon     & MTN Cameroon \\
China        & China Telecom, China Unicom \\
Egypt        & Orange, Etisalat \\
India        & Jio, Vi India, BSNL Mobile \\
Indonesia    & Indosat, 3, Smartfren \\
Iran         & TCI (Hamrah-e Avval), MTN Irancell \\
Japan        & au, SoftBank \\
Kenya        & Airtel \\
South Korea  & olleh (KT), LG U+ \\
Mexico       & AT\&T, Movistar \\
Morocco      & inwi, Orange Morocco \\
Nigeria      & Airtel, Glo \\
Pakistan     & Zong, Telenor, Ufone \\
Saudi Arabia & Mobily, Zain SA \\
Venezuela    & Digitel GSM, Movilnet \\
\bottomrule
\end{tabular}
\caption{Curated additional-majors override list (16 countries, 33
operators): operators treated as major (Definition~\ref{def:major}) although their OpenCellID tower counts fall below the 75\,\% threshold, sourced from national regulators and industry trackers. The complete per-country major-operator assignment (threshold-based majors plus overrides) ships with the data deposit.}
\label{tab:overrides}
\end{table}

\paragraph{Track B priors and constants.} The six free per-country
parameters carry literature-anchored priors: log-normal on the revenue-capture coefficients $\alpha_{c,t}$ ($\mu{=}0$, $\sigma{=}0.5$), on greenfield tower cost ($\mu{=}11.5$, $\sigma{=}0.4$; a \$80k--\$150k median) and on per-km fibre cost ($\mu{=}9.9$, $\sigma{=}0.4$; \$15k--\$40k/km), plus a $\mathrm{Beta}(2,38)$ prior ($\sim$5\,\% mean) on the share of revenue flowing to new-tower CapEx. Pooled-global constants held fixed during per-country fits are the upgrade cost (25\,\% of greenfield), a flat \$30k microwave-link cost, the footprint radii $r_t$, an 8\,\%-of-CapEx annual maintenance charge, and the maturity ramp $(1 - e^{-(y - y_0 + 1)/\tau_t})$ with $\tau_t \in \{3, 3, 5\}$ years for 2G/3G/4G. CMA-ES runs 200 evaluations at population size 12 and $\sigma_0{=}0.3$ in log-parameter space.

\begin{table}[!ht]
\centering
\footnotesize \setlength{\tabcolsep}{4pt}
\begin{tabular}{lll}
\toprule
\textbf{Track} & \textbf{Setting} & \textbf{Value} \\ \midrule
\multicolumn{3}{l}{\textit{Track A: LightGBM, one model per technology}}\\
 & objective / metric        & binary log-loss / log-loss\,+\,AUC \\
 & num\_leaves               & 63 \\
 & learning\_rate            & 0.03 \\
 & boosting rounds (early stop) & 2{,}000 (patience 100) \\
 & feature\_fraction         & 0.8 \\
 & bagging\_fraction / freq  & 0.8 / every 5 \\
 & min\_data\_in\_leaf       & 500 \\
 & $\ell_2$ leaf penalty     & 1.0 \\
 & per-country pixel cap     & $10^6$ per year \\
 & per-row sample weight     & $1/\sqrt{n_c}$ \\
 & calibration               & isotonic (held-out split) \\ \midrule
\multicolumn{3}{l}{\textit{Track B: structural simulator, CMA-ES per country}}\\
 & free params / country     & 6: $\alpha_{2/3/4\mathrm{G}}$, greenfield, fibre/km, CapEx share \\
 & priors                    & log-normal ($\alpha$, costs); $\mathrm{Beta}(2,38)$ (CapEx share) \\
 & ARPU disaggregation       & SEDAC GRDI multiplier, $\beta{=}0.3$ (fixed) \\
 & footprint radius $r_t$    & 12 / 7 / 5\,km (2G/3G/4G, rural) \\
 & maturity ramp $\tau_t$    & 3 / 3 / 5\,yr \\
 & upgrade / microwave cost  & 25\,\% of greenfield / \$30k flat \\
 & maintenance               & 8\,\% of CapEx per year \\
 & CMA-ES evals / pop / $\sigma_0$ & 200 / 12 / 0.3 (log space) \\
 & loss                      & weighted Dice on reliable triples \\
 & pooling ($<$3 triples)    & median posterior by WB income group \\
 & point estimate            & calibrated mean (deterministic simulator) \\ \midrule
\multicolumn{3}{l}{\textit{Track C: pix2pix U-Net, adversarial term disabled, one model per technology}}\\
 & input                     & 9-channel 256$\times$256\,km patch \\
 & loss weights (BCE/Dice)     & 1.0 / 0.5 \\
 & optimiser / learning rate & Adam / $10^{-4}$ \\
 & epochs / batch size       & 30 / 16 (Apple MPS) \\
 & augmentation              & flips $p{=}0.5$; $\pm 15^\circ$ rotation \\
 & uncertainty (internal)    & MC-dropout, $T{=}30$ \\
\bottomrule
\end{tabular}
\caption{Hyperparameter, calibration and uncertainty settings for the
three modelling tracks. Track~A and Track~C train one model per technology (2G/3G/4G); Track~B fits six free parameters per (country, technology) by CMA-ES, pooling sparsely-labelled cells to their World Bank income group. Internal track uncertainties feed validation only; the released interval derives from conformal calibration of the combined mean (Section~\ref{sec:bestestimate}).}
\label{tab:hyperparams}
\end{table}

\end{appendices}

\section*{Data availability}
The dataset is available from Zenodo at
\url{https://doi.org/10.5281/zenodo.21594337} under CC-BY-4.0. The rasters are
distributed as 22 ZIP archives, one per UN M49 subregion, each holding the full
1999--2030 series for every country in that subregion, for all four layers and
three technologies; \texttt{regions.csv} maps each country to its archive and a
manifest lists the size, file count, member countries and SHA-256 of every one. The reliability table
identifying which country--technology--year maps were used, the per-cell
launch-year certification and its source-level provenance, and the machine-
readable metadata are provided as separate files alongside the archives.

\section*{Acknowledgements}
This work received funding support from the Bill and Melinda Gates Foundation
(INV-045370) and supported by funding from the Alexander von Humboldt Foundation and the Federal Ministry of Education and Research (Bundesministerium f\"ur Bildung und Forschung) of Germany.

\bibliography{references}

\end{document}